\documentclass[12pt]{article}
\usepackage{amsfonts}

\usepackage{amsmath}
\usepackage{amssymb}
\csname @addtoreset\endcsname{equation}{section}
\newcommand{\be}{\begin{equation}}
\newcommand{\eeq}{\end{equation}}
\newcommand{\eqn}{\begin{eqnarray}}
\newcommand{\feqn}{\end{eqnarray}}
\newcommand{\arr}{\begin{eqnarray*}}
\newcommand{\farr}{\end{eqnarray*}}

\newcommand{\intM}{\int_{\cal M}}
\newcommand{\intdM}{\int_{\partial\cal M}}

\begin{document}

\begin{titlepage}
\begin{flushright}
UTHET-01-0801\\
hep-th/0108132
\end{flushright}
\vspace{.3cm}
\begin{center}
\renewcommand{\thefootnote}{\fnsymbol{footnote}}
{\Large \bf A brane in five-dimensional
Minkowski space\footnote{Research supported in
part by the DoE under grant DE-FG05-91ER40627.}}
\vfill
{\large \sc {J.~Avery,
R.~Mahurin and
G.~Siopsis}}\\
\renewcommand{\thefootnote}{\arabic{footnote}}
\setcounter{footnote}{0}
\vskip 1mm
{\normalsize
Department of Physics and Astronomy,\\
The University of Tennessee, Knoxville,
TN 37996 - 1200, USA.\\}
\end{center}
\vspace{2cm}

\centerline{\large \bf Abstract}
\vspace{.8cm}
\normalsize 

We discuss the propagation of gravity in five-dimensional Minkowski space in the
presence of a four-dimensional brane.
We show that there exists a solution to the wave equation that leads to
a propagator exhibiting four-dimensional behavior at low energies (long distances) with five-dimensional
effects showing up as corrections at high energies (short distances).
We compare our results with propagators derived in previous analyses exhibiting five-dimensional
behavior at low energies. We show that different solutions correspond to different physical systems.

\vfill

\end{titlepage}

Extra dimensions have been shown to lead to phenomenologically viable
results explaining the weakness of gravity compared to the other forces of
Nature. This has been achieved with both compact extra dimensions~\cite{bib1,bib2,bib3} and
uncompactified warped ones~\cite{bib4,bib5}. Typically, one obtains a tower of Kaluza-Klein
modes which contribute at high energies (short distances), whereas at
low energies (long distances) one recovers ordinary four-dimensional
gravity.

These features are not apparent when the extra dimensions are
flat~\cite{bib6,bib7,bib8}. It has been shown that four-dimensional gravity is not necessarily
recovered at low energies. This is due to the existence of light Kaluza-Klein
modes which may dominate even at low energies. It was shown in ref.~\cite{bib6}
that for one extra flat dimension one obtains five-dimensional
gravity at low energies, whereas in ref.~\cite{bib7} it was argued that for two
or more extra flat dimensions, low energy physics is dominated by
four-dimensional gravity. A more careful treatment of the singular limits
considered in ref.~\cite{bib8} revealed that the light Kaluza-Klein modes
dominate at low energies independently of the number of extra
flat dimensions.

In the analysis of ref.~\cite{bib6} the propagation of gravity in the extra dimension
was modeled by considering incoming waves
on both sides of the brane. If the extra dimension is given a finite size $R$, then
this approach corresponds to imposing Neumann boundary conditions on both ends of the
extra dimension~\cite{bib8}. Thus, one considers two distinct branes located at the ends of
the extra dimension. The infinite $R$ case is then recovered by letting the position of one brane go
to infinity. We shall instead identify the two ends, thus considering a single brane in a toroidally
compactified extra flat dimension. In this case, the Green function can be expressed in terms of
standing waves. We find that at low energies the propagator on the brane behaves as $1/p^2$,
where $p^\mu$ is the four-dimensional momentum,
i.e., as a four-dimensional propagator. This is our main result.
We should point out that it is not surprising that different boundary conditions lead to distinct
physical systems. This is because, unlike in AdS space (see, e.g.,~\cite{bib9}), there is no holomorphic
principle in Minkowski space.

Following refs.~\cite{bib6,bib7,bib8} we discuss the propagation of a single scalar field
$\phi$ instead
of a graviton. This simplifies the discussion because the complications due to
the tensor structure of the graviton are avoided. The field $\phi$ is a function of five coordinates $x^I = (x^\mu , y)$ where $x^\mu$ ($\mu = 0,1,2,3$) span
our four-dimensional world and $y$ is the coordinate of the extra dimension.
We will first consider a finite extra dimension of size $R$ (so that $0\le y\le R$) and then study the
large $R$ limit.
The brane is located at the four-dimensional slice $y=0$.
The action is
\be S = M^3 \intM d^4x dy \partial_I\phi \partial^I\phi + \bar M^2 \intdM
d^4x \partial_\mu \phi \partial^\mu\phi
\end{equation}
where $M$ ($\bar M$) is the five (four) dimensional mass scale.
The manifold ${\cal M}$ is a portion of the five-dimensional Minkowski space
bounded by the brane.
The field equation one derives from this action is
\be\label{eq11} \Big[ M^3 (\partial_\mu\partial^\mu + \partial_y^2) + \bar M^2 \delta (y)\partial_\mu\partial^\mu \Big] \phi (x^\mu, y) = 0
\end{equation}
Taking fourier transforms in the four-dimensional space,
\be \phi (x^\mu, y) = e^{ip\cdot x} f(y)
\end{equation}
where $p_\mu = (\omega, \vec p)$ and we adopt a mostly positive signature
for the Minkowski space,
the wave equation becomes
an ordinary differential equation
\be\label{eq5} M^3 f''(y) - (M^3 + \bar M^2 \delta (y))p^2 f(y) = 0
\end{equation}
to be solved in the domain of $y$ which will be restricted by the boundary defined by the
brane, i.e., $(0,\infty)$.
The bulk-to-brane Green function satisfies the equation
\be\label{eq2} M^3 G''(p;y,0) - (M^3 + \bar M^2 \delta (y))p^2 G(p;y,0) = \delta(y)
\end{equation}
where we are using the notation $G(p;y,y')$ for the Green function.
As was shown in ref.~\cite{bib6}, this can be expressed in terms of the Green function in the
absence of the brane, satisfying the equation
\be\label{eq2a} D''(p;y,0) - p^2 D(p;y,0) = i\delta(y)
\end{equation}
where we included a factor of $i$ in the $\delta$-function for convenience.
The associated eigenvalue problem is
\be f'' (y) +q^2 f(y) = \lambda f(y) \end{equation}
where $q=\sqrt{-p^2} = \sqrt{\omega^2 -\vec p^2}$.
The eigenfunctions $f(y)$ obey Neumann boundary conditions at the end-points $y=0,R$,
\be f_n (y) = \cos {n\pi y\over R}\quad,\quad \lambda_n = q^2 - {n^2\pi^2\over R^2}\end{equation}
Therefore, we may expand~\cite{bib8}
\be\label{dexp} D(p;y,0) = {i\over R} \; \sum_{n=1}^\infty {f_n(y)\over \lambda_n} =
{i\over 2R} \; \sum_{n=-\infty}^\infty {e^{in\pi y/R}\over \lambda_n}\end{equation}
In the large $R$ limit ($R\to\infty$), the sum becomes an integral,
\be\label{dexp2} D(p;y,0) =
\int_{-\infty}^\infty {dz\over 2\pi i}\, {e^{iyz}\over z^2 -  q^2}\end{equation}
Since we wish to propagate disturbances from the point $y$ in the bulk to the brane at $y=0$, we
need to adopt the prescription $q^2 \to (q+i\epsilon)^2$, which selects the pole at $z=q$. We obtain
\be\label{dexp3} D(p;y,0) = {1\over 2q}\; e^{iqy} \end{equation}
The original Green function $G(p;y,0)$ is proportional to $D(p;y,0)$. Setting
$G(p;y,0) = A\, D(p;y,0)$, and plugging into eq.~(\ref{eq2}), we obtain
\be \label{eq10a} G (p; y,0) = {1\over iM^3 + \bar M^2 q^2 D(p;0,0)} \; D(p;y,0)\end{equation}
In the large $R$ limit, this becomes
\be \label{eq10} G (p; y,0) = {1\over 2i M^3 q + \bar M^2 q^2} \; e^{iqy}\end{equation}
leading to five-dimensional behavior, $G\sim 1/q$, on the brane for small
momenta. It follows that the Kaluza-Klein modes dominate at low energies.

Next, we show that the wave equation possesses a different solution behaving differently at low
energies. This is possible, because, unlike AdS space (see, e.g., ref.~\cite{bib9}), Minkowski space
does not satisfy any holographic conditions. Of course, a Green function that behaves qualitatively
differently at low energies must correspond to a different physical system. We shall first introduce
the new solution and then study its properties. It is not hard to see that in the large $R$ limit, 
the Green function
\be\label{eq1} G(p; y,0) = {1\over \bar M^2 q^2}\; \cos (qy)\end{equation}
satisfies eq.~(\ref{eq2}) and Neumann boundary conditions on the brane. Thus, in this case the Kaluza-Klein modes
decouple at low energies and we obtain four-dimensional behavior of the propagator on the brane.
For finite $R$, the analogue of this Green function can be obtained by identifying the two ends of the
extra $y$-dimension (toroidal compactification). Then disturbances can propagate in both directions
along $y$. This implies that we need to include the contributions of both poles in the definition of
the auxiliary Green function $D(p;y,0)$ (Eqs.~(\ref{dexp}), (\ref{dexp2})), or equivalently, define the
new Green function in terms of the real part of $D(p;y,0)$. Then the disturbance originating from
$y$ may reach the brane at $y=0$ by traveling to the left as well as to the right (toward the point $y=R$, which is identified with the $y=0$ end). Thus, we obtain standing waves. Setting $G(p;y,0) = A\; \Re D(p;y,0)$, and plugging in
eq.~(\ref{eq2}), we obtain
\be\label{eq1a} G(p; y,0) = - {2\over \bar M^2 q}\; \Re D(p;y,0)\end{equation}
which reduces to eq.~(\ref{eq1}) in the large $R$ limit.

The reader may object that the above result is unreliable, because it is a solution to a singular limit,
eq.~(\ref{eq2}). Indeed, we have idealized the brane as a lower dimensional hypersurface in the
five-dimensional Minkowski space. A physical brane is expected to have a small but finite width in
the $y$-direction
of order $1/M$, where $M$ is the fundamental scale of the underlying higher-dimensional theory.
A proper derivation of the Green function would rely upon a regularization of the brane action.
This need for regularization was recognized in ref.~\cite{bib8} where a cutoff was
introduced for massive Kaluza-Klein modes.
This did not alter the low energy five-dimensional behavior of the propagator.

We shall now show
that the introduction of a small but finite width of the brane does not alter the main conclusion~(\ref{eq1}), i.e., that low energy physics is four-dimensional if we identify the ends
$y=0$ and $y=R$, thus having one brane in a toroidally compactified extra dimension.
Our regulator will introduce
the effects of the Kaluza-Klein modes and it will be shown that their effects amount to high-energy corrections
of order $1/M$. We shall calculate these effects in the large $R$ limit where explicit results can be derived.

We shall get rid of the singularity in the position of the brane by giving it
a small but finite extent $1/\Lambda$ in the $y$-direction, where $\Lambda
\sim M$.
Thus, we shall approximate the $\delta$-function by
\be \delta (y) \approx V(y) = \left\{ {\Lambda e^{\Lambda y} \quad,\quad y \le 0\atop 0\quad,\quad y>0}\right.
\end{equation}
and instead of eq.~(\ref{eq5}) consider the differential equation
\be M^3 f''(y) - (M^3 + \bar M^2 V (y))p^2 f(y) = 0
\end{equation}
in the domain $(-\infty,\infty)$, keeping in mind that this is the large $R$ limit of a finite extra dimension. The brane now occupies the negative $y$-axis, whereas the
``bulk'' is still the positive $y$-axis and $y=0$ is its boundary defined by the brane.
The corresponding Green function satisfies
\be M^3 G''(p;y,y') - (M^3 + \bar M^2 V (y))p^2 G(p;y,y') = \delta (y-y')
\end{equation}
where we differentiate with respect to the first argument, $y$.
In the bulk ($y>0$), the wave equation
admits plane-wave solutions,
\be f(y) \sim e^{\pm ipy}\end{equation}
Since we are interested in standing waves, 
we shall consider incoming waves, $f(y)\sim e^{ipy}$ and at the end extract the real part of the
resulting Green function ({\em cf.}~eq.~(\ref{eq1a})). In the brane ($y<0$), the wave equation admits
solutions
\be f(y) \sim Z_\nu (\beta e^{\Lambda y/2})\quad,\quad \nu = 2iq/\Lambda\quad,\quad
\beta = \sqrt{-4p^2\bar M^2\over M^3\Lambda}\end{equation}
where $Z_\nu$ is a Bessel function. As $y\to -\infty$, we want the solution to behave as an outgoing
wave. This implies that $f(y) \sim J_\nu$ in the region that includes $-\infty$
(since $J_\nu (\beta e^{\Lambda y/2} ) \sim e^{\nu\Lambda y/2} = e^{iqy}$).

The Green function satisfies the wave
equation if $y\ne y'$. Let us first calculate the bulk-to-brane Green function by placing $y'$
on the brane ($y'<0$).
For $y<y'$, the Green function is
\be G(p;y,y') = G_< (p;y,y') = A(y') J_\nu (\beta e^{\Lambda y/2})\end{equation}
where we kept the solution that approaches an outgoing wave as $y\to -\infty$.

The interval $y>y'$ is split into two intervals, inside the brane, $(y',0]$ and  in the bulk, $(0,\infty)$. For $y>0$,
the Green function behaves as an incoming plane wave,
\be\label{eq15} G(p;y,y') = G_> (p;y,y') =  A'(y') e^{iqy}\end{equation}
whereas in the brane ($y\le 0$), the solution is a linear combination of Bessel functions,
\be G_> (p;y,y') = B(y') J_\nu (\beta e^{\Lambda y/2} ) + C(y') Y_\nu (\beta e^{\Lambda y/2} )\end{equation}
Matching the two forms of $G_> (p;y,y')$ at the boundary point $y=0$ leads to
\be A'(y') = - {2\over\pi\beta Y_{1+\nu} (\beta)}\; B(y') \quad,\quad C(y') = - {J_{1+\nu} (\beta) \over Y_{1+\nu } (\beta)} \; B(y') \end{equation}
where we used the recursion relation for Bessel functions,
$zZ_\nu' (z)-\nu Z_\nu (z) +zZ_{1+\nu } (z) =0$.
We need to match the two expressions $G_>$ and $G_<$ at $y=y'$.
\be G_> (p;y,y')\Big|_{y=y'} = G_< (p;y,y') \Big|_{y=y'}\end{equation}
\be M^3 (\partial_y G_> (p;y,y') - \partial_y G_< (p;y,y') ) \Big|_{y=y'} = 1\end{equation}
We obtain
\be\label{eq20} A'(y') = {2\over M^3\Lambda \beta J_{1+\nu } (\beta)} \; J_\nu (\beta e^{\Lambda y'/2} )\end{equation}
Combining~(\ref{eq15}) and (\ref{eq20}), we obtain
the bulk-to-brane Green function as the real part of
\be\label{eq6} G(p;y,y') = {2\over M^3\Lambda \beta J_{1+\nu } (\beta)} \; J_\nu (\beta e^{\Lambda y'/2} ) e^{iqy} \quad,\quad y>0\quad,\quad y'<0\end{equation}
To separate the effects of the Kaluza-Klein tower from the zero modes, we use the Bessel
function identity~({\em cf.}~ref.~\cite{bib10})
\be Z_\nu (z) = {2(1+\nu )\over z}\; Z_{1+\nu } - Z_{2+\nu}\end{equation}
Eq.~(\ref{eq6}) becomes
\be G(p;y,y') = {4(1+\nu)\over M^3\Lambda\beta^2}\; e^{iq(y-y')}
- {2\over M^3\Lambda \beta J_{1+\nu } (\beta)} \; J_{2+\nu } (\beta e^{\Lambda y'/2} ) e^{iqy}\end{equation}
The first term is the contribution of the zero mode whereas the second term
is due to the Kaluza-Klein tower.
In the small momentum limit, we may approximate $J_\nu (\beta) \approx \beta^\nu /\Gamma (1+\nu )$.
The Green function becomes the real part of
\be G(p;y,y') \approx - {1\over \bar M^2 p^2} \; e^{iq(y-y')}\end{equation}
and the propagator on the brane is
\be G(p;0,0) \approx - {1\over \bar M^2 p^2} \end{equation}
Thus the propagator behaves as a four-dimensional propagator at low energies
and the Kaluza-Klein modes decouple, in agreement with our earlier result~(\ref{eq1}).

For comparison with the result of ref.~\cite{bib6} (eq.~(\ref{eq10})), 
let us assume that the two ends $y\to \pm\infty$ are not to be identified. We then require the Green function to behave as $G\sim J_{-\nu}$ in the region that includes $y\to -\infty$, so that a disturbance
originating in that region will propagate toward $y=0$ and not away from the brane. By following a similar argument
as before, we obtain the bulk-to-brane propagator
\be G (p;y,0) = {2\over M^3\Lambda \beta } \; {J_{-\nu} (\beta)\over
J_{1-\nu } + {2\nu\over\beta}\; J_{-\nu} (\beta)} \; e^{iqy}\end{equation}
to be contrasted to eq.~(\ref{eq6}).
The small momentum limit ($\Lambda\to\infty$) of this expression is
\be G (p;y,0) \approx {1\over \Lambda M^3 (\beta^2/4 + \nu)}
e^{iqy} = {1\over \bar M^2 q^2 + 2iM^3 q} \; e^{iqy} \end{equation}
in agreement with eq.~(\ref{eq10}).

In conclusion, 
we have shown that it is possible to obtain four-dimensional dynamics on a
brane embedded in a five-dimensional Minkowski space.
The difference between our approach and previous analyses~\cite{bib6,bib7,bib8}
is in the treatment of the end points of the extra dimension.
By assuming that two distinct branes lie on the two ends of the extra dimension,
one obtains five-dimensional low energy behavior of the propagator~(\ref{eq10}).
If, instead, the two ends are identified (toroidal compactification), thus having only one brane
in the system, we recover four-dimensional low energy dynamics
and the Kaluza-Klein modes decouple ({\em cf.}~eq.~(\ref{eq1})).
Obtaining different physical systems by choosing different boundary conditions is possible
in Minkowski space because there is no holographic principle similar to AdS space~\cite{bib9,bib10}.
These
results persist when the brane is assumed to have a finite thickness in the
extra dimension, as is physically expected.
It would be interesting to generalize the discussion to higher-dimensional
Minkowski spaces and consider branes of co-dimension larger than 1.
Work in this direction is in progress.



\end{document}